\documentclass[
twocolumn,
]{ceurart}

\usepackage{listings}
\begin{document}

\copyrightyear{2022}
\copyrightclause{Copyright for this paper by its authors.
  Use permitted under Creative Commons License Attribution 4.0
  International (CC BY 4.0).}

\conference{IJCAI 2023 Workshop on Deepfake Audio Detection and Analysis (DADA 2023), August 19, 2023, Macao, S.A.R}

\title{An End-to-End Multi-Module Audio Deepfake Generation System for ADD Challenge 2023}


\author[1]{Sheng Zhao}[%
orcid=0000-0001-9376-1615,
email=zhaosheng@lyxxkj.com.cn,
]

\fnmark[1]
\address[1]{NanJing LongYuan Information Technology Co.Ltd,NanJing,China}

\author[1,2]{Qilong Yuan}[%
orcid = 0009-0009-5910-7957,
email=yuanqilong@lyxxkj.com.cn,
]
\fnmark[1]
\address[2]{College of Computer and Information,Hohai University,NanJing,China}

\author[1]{Yibo Duan}[%
orcid = 0000-0001-9512-3588,
email=duanyibo@lyxxkj.com.cn,
]

\author[1]{Zhuoyue Chen} [%
orcid =0009-0007-9223-8962,
email=chenzhuoyue@lyxxkj.com.cn,
]

\cormark[1]
\cortext[1]{Corresponding author.} 
\fntext[1]{These authors contributed equally.}

\begin{abstract}
  The task of synthetic speech generation is to generate language content from a given text, then simulating fake human voice. 
  The key factors that determine the effect of synthetic speech generation mainly include speed of generation, accuracy of word segmentation, 
  naturalness of synthesized speech, etc. This paper builds an end-to-end multi-module synthetic speech generation model, including speaker encoder, 
  synthesizer based on Tacotron2, and vocoder based on WaveRNN. 
  In addition, we perform a lot of comparative experiments on different datasets and various model structures. Finally, 
  we won the first place in the ADD 2023 challenge Track 1.1 with the weighted deception success rate (WDSR) of 44.97\%.
\end{abstract}


\maketitle
\section{Introduction}

The second Audio Deepfake Detection Challenge (ADD 2023) \cite{yi2023add} aims to spur researchers 
around the world to build new innovative technologies that can further accelerate and foster research on detecting and analyzing deepfake speech utterances. 
Among them, the task of Track 1.1 is to generate fake audio from some given text. 
At the same time, the generated fake audio from Track 1.1 is detected according to the detection model in Track 1.2 and the baseline RawNet2 \cite{9414234}. 
In this paper, we propose an end-to-end audio generation system. 
The system, consisting of three parts, speaker encoder, synthesizer and vocoder, uses neural network models to directly convert text into speech signals. 
At present, speech synthesis technology is in a stage of rapid development, 
and many cutting-edge technologies are constantly emerging.
\begin{enumerate}
  \item With the development of deep learning, end-to-end models have emerged, such as Tacotron \cite{wang2017tacotron}, 
  which consists of an encoder and a decoder. The encoder is responsible for converting the input text into feature representation, 
  and the decoder generates the corresponding speech signal from these features. 
  The decoder models different parts of the input more finely based on the attention mechanism, improving the quality and naturalness of the synthesized speech. 
  \item Speech synthesis based on autoregressive model, such as WaveNet \cite{oord2016wavenet}, WaveRNN \cite{kalchbrenner2018efficient}. 
  By controlling the input text, the speech samples of each time step are sequentially generated, and a complete speech signal is synthesized. 
  \item Speech synthesis techniques based on Generative Adversarial Networks(GAN) \cite{goodfellow2014generative}, 
  such as HiFi-GAN \cite{kong2020hifigan} and Fre-GAN \cite{kim2021fregan}, pit the generator and the discriminator against each other, gradually improving the quality and fidelity of the generated speech.
\end{enumerate} 
Nevertheless, traditional speech synthesis technology has shortcomings such as large data requirements, complicated training process, and unnatural sound quality. 
Inspired by SV2TTS \cite{jia2019transfer}, this paper proposes an end-to-end autoregressive fake audio generation model. 
Specifically, the model is divided into three parts: speaker encoder, synthesizer and vocoder. 
Speaker encoder adopts a speech coding network constructed by BiLSTM \cite{huang2015bidirectional} and a fully connected layer. 
The network provides speaker classification information for the synthesizer and realizes multi-speaker speech synthesis. 
Synthesizer adopts the architecture of Tacotron2 \cite{8461368}, which can generate speech features with high quality. 
Vocoder adopts WaveRNN, one of the autoregressive model structures, which can gradually learn the change rule of the input Waveform and generate high-fidelity output Waveform.

The rest of this paper is organized as follows: Section 2 describes our proposed method in detail. 
See Section 3 for the experimental results and Section 4 for the conclusion.



\section{System Description}
\begin{figure*}[htbp]
\centering 
\includegraphics[width=\linewidth]{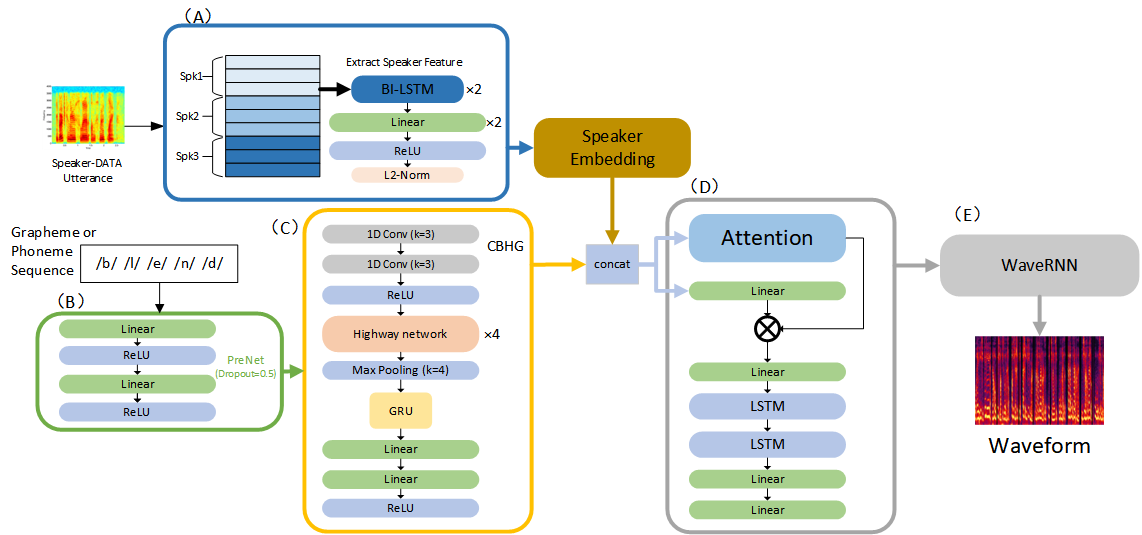}
\caption{\centering{Model overview. Each of the three components is trained independently.}} 
\label{system_framework}
\end{figure*}

In this section, we will first introduce the framework of our system, and then describe the reasons for choosing it to generate fake speech. 
Our system framework is shown in Figure \ref{system_framework}. It mainly consists of speaker encoder, synthesizer based on Tacotron 2, and vocoder based on WaveRNN. 
Subsequently, we will introduce the objective function used in training each of our modules.

\subsection{Speaker Encoder} 
Specifically, two layers of bi-LSTM are used to capture the temporal dependencies and context information of the input speech sequence. 
Bi-LSTM can effectively learn variable-length sequence data and produce encoded intermediate speech feature representations. 
Two fully connected layers further enhance the abstraction of speech features and obtain low-dimensional speech embeddings. 
The fully connected layer can greatly compress the speech feature dimensions while retaining semantic information, 
producing highly abstract speech embedding representations.

To eliminate the dimensional influence of speech embedding vectors and make the model focus on direction rather than length, 
speech embeddings are L2 normalized. Normalization can make cosine similarity calculation more accurate, 
thereby improving the performance of speech similarity judgment.

This speech encoder can learn speech embeddings that express rich speaker-related information, 
providing speaker classification information for the Tacotron 2 framework to achieve multi-speaker speech synthesis. 
Tacotron 2 alone is difficult to distinguish speech features of different speakers. 
Combining this speech encoder can effectively solve this problem, achieving high-quality cross-speaker speech synthesis.
Therefore, the speaker encoder we use can learn highly abstract speech embedding representations and encode rich speaker information. 
Through training, this model can learn the semantic information of speech and efficiently complete speaker classification, 
providing valuable speech expression methods for speech-related research.

\subsection{Synthesizer} 
Synthesizer adopts the Tacotron 2 architecture, consisting of a PreNet(Double fully connected layer), CBHG(1-D Convolution Bank + Highway network \cite{srivastava2015highway} + bidirectional GRU) module, 
attention mechanism and decoder network (as shown in modules B, C and D in Figure \ref{system_framework}). 
It is a powerful sequence-to-sequence model that can generate high-quality speech features.

The PreNet and CBHG module together constitute the encoder, 
which can learn the high-level feature representation of the Mel spectrogram sequence and jointly encode it. 
The encoder provides the attention mechanism with ideal conditional information to achieve complete and accurate attention alignment.
The CBHG module consists of a one-dimensional convolution layer, highway network and bidirectional GRU. 
It can learn the high-level feature representation and nonlinear dependencies of the Mel spectrogram. 
The output of the CBHG module provides the attention mechanism with accurate conditional information, which has an important influence on its performance.
The attention mechanism realizes dynamic speech feature generation at each time step. 
It can learn the alignment relationship between the PreNet and CBHG module outputs to produce attention weights. 
The attention weights determine the content of the speech features generated at each time step. 
The PreNet can further improve the accuracy of speech feature prediction. 
It can eliminate the influence of the previous time step prediction, making the prediction at each time step more independent.
The bidirectional LSTM can learn the historical and future information of the speech feature sequence jointly, 
generating an accurate prediction at each time step. It can achieve continuous and smooth speech feature generation.

Synthesizer is a key component for achieving multi-speaker speech synthesis, 
providing a powerful speech feature generation module for the complete speech synthesis model.

\subsection{Vocoder}
The vocoder adopts the WaveRNN structure, which can learn high-dimensional conditional information of speech and generate highly realistic speech waveforms. 
WaveRNN is a recurrent neural network model that can gradually learn the changing rules of the input waveform and produce high-fidelity output waveforms.
WaveRNN has the following advantages:
\begin{enumerate}
\item It can learn complex high-dimensional data distributions; 
\item It has a memory mechanism and can learn long-term dependency relationships;
\item It is easy to train and fast convergence.
\end{enumerate} 
This model can generate high-fidelity speech waveforms and provide powerful speech generation capabilities for speech synthesis systems.

\subsection{Training Loss}
For the speaker encoder, we adopt the cross-entropy loss\cite{mao2023crossentropy}, 
which can enable the speaker encoder to learn speech embeddings that distinguish between different speakers. 
The cross-entropy loss can measure the difference between the predicted distribution and the true distribution, 
so that the model parameters are updated in the direction of reducing this difference. 
Therefore, The formula for speaker classification loss $L_{speaker}$ is
\begin{align}
  L_{speaker} = - \sum_{i=1}^{N} y_{i} \log p_{i}
\end{align}
where $N$ represents the number of different speakers in each batch of training data. 
the $y_i$ is the one-hot encoded target speaker, 
and $p_i$ is the predicted speaker probability distribution of the model.

For synthesis loss, we adopt L1 loss which can enable the synthesizer to learn to generate output Mel-spectrum close to the target one.
It can directly measure the difference between the predicted value and the target value, so that the model learns to minimize this difference. 
The L1 loss with periodic mask can enable the synthesizer to focus on the fine structure of the Mel-spectrum within the same pitch period. 
The combination of these two loss functions can enable the synthesizer to generate high-quality Mel-spectrum sequences. The formula for synthesis loss $L_{synthesis}$ is:
\begin{align}
L_{synthesis} = |mel_{target} - mel_{predict}|
\end{align}
where $mel_{target}$ is the target Mel-spectrum, and $mel_{predict}$ is the predicted Mel-spectrum of the model.
The formula for L1 loss with periodic mask $L_{cyc}$ is:
\begin{align}
L_{cyc} = |mel_{target} - mel_{predict}| * M
\end{align}
where $M$ is the periodic mask that can enhance the loss within the same pitch period.

For the vocoder loss, firstly we quantize the audio waveform into discrete values, and then train the model using the cross-entropy loss, which can measure the gap between the probability distribution predicted by the model and the true label.
Therefore, The formula for vocoder loss $L_{vocoder}$ is
\begin{align}
  L_{vocoder} = - \sum_{i=1}^{M} g_{i} \log p_{i}
\end{align}
where $M$ represents the discrete-valued dimensions for audio waveform quantization,
the $g_i$ is the gap between the predicted value and the true label, 
and $p_i$ is the probability distribution predicted by our model.

In summary, this model contains three modules: the speech encoder, the synthesizer and the vocoder. 
The three modules correspond to learning high-level speech semantics, intermediate speech features and low-level speech waveforms respectively. 
This model combines deep neural networks with waveform generation. 
It can both simulate the speech waveform and the internal features of the speech waveform, 
preserving richer speech information and generating highly realistic multi-speaker speech.

\section{Experiments}
\label{sec:system_framework}
\subsection{Dataset} 
We use the AISHELL3 \cite{shi2021aishell3} dataset to train the three modules of our model: 
the synthesizer, the speaker encoder and the vocoder.
AISHELL-3 is a large-scale open-source Chinese speech dataset. It contains over 300,000 speech utterances . 
The speech samples are recorded at 16KHz with 16bit quantization, and the duration of each utterance is 5 to 15 seconds.
For the speaker encoder, we use data augmentation methods to improve its generalization ability. 
We add noise(from MUSAN dataset \cite{musan2015}), reverberation(from RIRs dataset\cite{7953152})  
and speed perturbation on the training speech. 
These data augmentation methods can generate new training samples without changing the speech content, 
enrich the model's training data and enhance the generalization of the model.


\subsection{Comparison of different methods}
To verify the effectiveness of our proposed model, we conduct comparative experiments on different speech generation models and various datasets. 
The datasets used include AISHELL3 and LibriTTS, and the speech generation models are fastspeech \cite{ren2019fastspeech} and the model proposed in this paper. 
We construct two datasets for testing, the one is generated by two models on AISHELL3, and the other is generated by two models on LibriTTS\cite{zen2019libritts}. Both of them consist of 500 fake audios and 500 real audios.
We also select three synthetic speech detection models to calculate the EER \cite{delgado2021asvspoof} of the test set, the detection models are RawNet2, Res-TSSDNet\cite{9456037} and ECAPA-TDNN\cite{das2021hlt}. 
The RawNet2 model uses the pretrained model of ASVSpoof2021.Res-TSSDNet and ECAPA-TDNN are trained on the data from ADD2023 track 1.2. The relevant results are shown in Table \ref{table1}. 
We can see that our proposed model is superior to the baseline model of fastspeech, and the EER on the LibriTTS test set reaches 58.71\%. 
In addition, on some detection models, the EER of the proposed model is more than twice that of the comparison model.

\begin{table}
  \centering
  \caption{EERs of different generative models and detection models.}
  \label{table1}
  \resizebox{0.9\linewidth}{!}{%
  \begin{tabular}{cccc}
  \toprule
  \textbf{Detection model} & \textbf{Dataset}   & \textbf{Method} & \textbf{EER(\%)}\\ 
  \midrule
  RawNet2& LibriTTS & Proposed & 58.71\\
      &  & Fastspeech & 22.02\\
     &   AISHELL3  & Proposed & 21.84\\
     &  & Fastspeech & 8.02\\
  \midrule
  Res-TSSDNet & LibriTTS & Proposed & 39.17\\
  &  & Fastspeech & 23.39\\
  &AISHELL3 & Proposed & 56.71\\
  &  & Fastspeech & 16.54\\
  \midrule
  ECAPA-TDNN & LibriTTS & Proposed & 61.12\\ 
  &  & Fastspeech & 60.62\\
  &AISHELL3 & Proposed & 63.52\\
  &  & Fastspeech & 61.74\\
  \bottomrule
  \end{tabular}}
\end{table}

Moreover, in order to verify the influence of the internal structure of the model on the generated audio, we also conduct corresponding ablation experiments. 
Specifically, we replace WaveRNN of the proposed vocoder model with Hifi-GAN to conduct EER tests on the two datasets. 
The results are shown in Table \ref{table2}, and it can be seen that the effect of using WaveRNN as a vocoder is better than HifiGAN on both datasets. 
On LibriTTS, using Hifi-GAN results in about 7\% EER drop, and on AISHELL3, 15\% EER drop.

\begin{table}
  \centering
  \caption{EERs with different vocoder structures.}
  \label{table2}
  \resizebox{0.9\linewidth}{!}{%
  \begin{tabular}{cccc}
  \toprule
  \textbf{Detection model} & \textbf{Dataset}   & \textbf{Vocoder} & \textbf{EER(\%)}\\ 
  \midrule
  RawNet2& LibriTTS & WaveRNN & 58.71\\
      &  & Hifi-GAN & 51.50\\
     &   AISHELL3  & WaveRNN & 21.84\\
     &  & Hifi-GAN & 6.41\\
  \bottomrule
  \end{tabular}}
\end{table}

In order to improve the authenticity and similarity of the audio generated by the model, 
we stitch all the audio of the specific speaker as the voice input file of the model.
we also conduct relevant contrast experiments to verify the effect of not performing audio stitching on the results. 
One set of generated audios is the result of the concatenated audio of the speaker as the voice line input, 
while the other set is the result of the single audio of the speaker as the voice line input, 
and the corresponding EER is calculated respectively, and the results are shown in Table \ref{table3}. 
The results show that whether the audio in AISHELL3 or LibriTTS is used as input, 
the results after audio splicing are better than the results before audio splicing.

\begin{table}
  \centering
  \caption{EERs with different input methods.}
  \label{table3}
  \resizebox{0.9\linewidth}{!}
  {\begin{tabular}{cccc}
    \toprule
    \textbf{Detection model} & \textbf{Dataset}   & \textbf{Splicing audio} & \textbf{Unspliced audio}\\ 
    &  &  EER(\%) & EER(\%) \\
    \midrule
    \large RawNet2& \large LibriTTS & \large 58.71 & \large 41.21\\
    & \large AISHELL3 & \large 21.84 & \large 17.84\\
    \bottomrule
  \end{tabular}}
\end{table}

\subsection{Evaluation}
Track 1.1 requires teams to generate attack samples based on given text and speaker identity. 
When testing the quality of synthesized audio, we include real voices of the corresponding speakers in the dataset. 
This allows for better model evaluation through EER. 
The official competition uses all Track 1.2 detection models as a confrontation, 
and finally uses the deception success rate(DSR) for ranking, and can also evaluate the effectiveness of the model.

\begin{figure}[htbp]
  \centering{
  \includegraphics[width=0.5\textwidth]{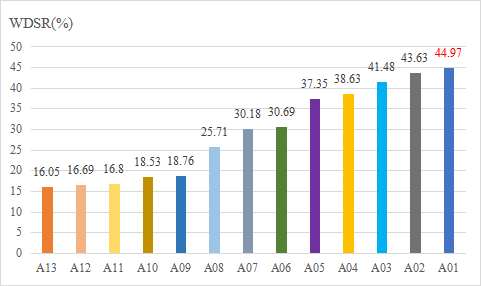}
    }
  \caption{\centering{Weighted deception success rate(WDSR) team rank. }}
  \label{track1}
\end{figure}

Weighted deception success rates(WDSR) of each team are shown in Figure \ref{track1}. 
We won the first place among all the participating teams. 
This also proves the rationality and effectiveness of our proposed methods and experiments.

\section{Conclusion}
In this paper, we propose an end-to-end multi-module synthetic speech generation model. 
In addition, we have done a lot of comparative experiments on different datasets and model structures, 
which proves that our model is logical and effective. The model ranked first in the ADD 2023 Challenge .

\section{Acknowledgement}
We would like to express our sincere gratitude to all those who helped and supported us during the writing of this paper and development of the system. 
First, we would like to thank the organizers for hosting the ADD 2023 Challenge. 
The Challenge provided us a platform to test and improve our technology and capabilities in voice synthesis generation. 
We won the first prize in the challenge with our state-of-the-art system design. 
Moreover, we would especially like to thank our company Nanjing Longyuan Information Technology Co., Ltd. 
The computing resources provided by our company offered us the opportunity to develop the most advanced deep neural networks on large datasets.

\bibliography{lyrference}

\end{document}